\documentclass[twocolumn,amsmath,amssymb,superscriptaddress]{revtex4}
\usepackage{graphicx}
\usepackage{dcolumn}
\usepackage{bm}
\def\bea {\begin{eqnarray}}
\def\eea {\end{eqnarray}}

\def\be {\begin{equation}}
\def\ee {\end{equation}}

\begin{document}
\title{Finite size effect on thermodynamics of hadron gas in high-multiplicity events of proton-proton collisions at the LHC}
\author{\it Nachiketa Sarkar}
\address{Variable Energy Cyclotron Centre, HBNI, 1/AF Bidhan Nagar, Kolkata 700 064, India}
\author{\it Paramita Deb}
\address{Indian Institute of Technology Bombay, Mumbai 400 076, India}
\author{\it Premomoy Ghosh}
\email{prem@vecc.gov.in}
\address{Variable Energy Cyclotron Centre, HBNI, 1/AF Bidhan Nagar, Kolkata 700 064, India}
\date{\today}         
\begin{abstract}
Multiple Reflection Expansion (MRE) formalism has been applied to hadron resonance gas (HRG) model to study the finite-size effect on thermodynamics of small systems of hadron gas at the chemical freeze-out temperature in high-multiplicity events of proton-proton (pp) colisions at the LHC. Comparison with larger systems of heavy-ion (AA) collisions helps in undersanding the usefulness of the effect on small systems. Thermodynamic properties of these systems at the chemical freeze-out, with and without system-size effect, are contrasted with those for infinite hadronic phase of strongly interacting matter at ideal thermodynamic limit, as provided by LQCD calculations. On introduction of finite size effect, the small hadronic systems produced in high-multiplicity pp events, unlike those in AA collisions, remain away from ideal thermodynamic limit. Knudsen number estimations validate the findings. \\
\end{abstract}
\maketitle
\section{Introduction}
\label{}
Quark-gluon plasma (QGP) \cite {ref01, ref02}, the thermalized partonic phase of strongly interacting matter that is governed by the quantum chromodynamics (QCD), is formed in the laboratory in relativistic collisions of heavy-ions (AA), like AuAu at center-of-mass energy, $\sqrt s_{NN}$ = 130 \& 200 GeV \cite {ref03, ref04, ref05, ref06} at the Relativistic Heavy-ion Collider (RHIC) and PbPb at $\sqrt s_{NN}$ = 2.76 \& 5.02 TeV \cite {ref07, ref08, ref09} at the Large Hadron Collider (LHC). The $\sqrt s_{NN}$ of collisions at RHIC and LHC correspond to small or vanishing baryon chemical potential ($\mu_{B}$) at high temperature ($T$) in the QCD phase diagram. The first-principle Lattice QCD (LQCD) calculations at vanishing $\mu_{B}$ predict crossover \cite {ref10, ref11} between the partonic and the hadronic QCD-phases. The present understanding on particle production in ultra-relativistic AA collisions in the vanishing $\mu_{B}$ region visualize formation of a fireball of de-confined partons due to high initial energy-density imparted by the colliding ions. Collisions among de-confined partons, in the initial stage, give rise to a system of expanding partonic medium in local thermal equilibrium, the QGP, that eventually undergoes a crossover to hadronic phase. The hadronic phase, in thermal and chemical equilibrium, reaches chemical freeze-out when inelastic collisions among the hadrons cease, freezing the abundances of final-state hadrons. Subsequently, the elastic collisions also are stopped and the kinetic freeze out occurs. \\
The hadron resonance gas (HRG) model \cite {ref12}, by matching the measured final-state hadron-yields, helps in finding the chemical freeze-out parameters, the baryon chemical potential ($\mu_{B}^{ch}$), the temperature ($T^{ch}$) and the volume ($V^{ch}$) of the equilibrated hadron gas, considered to have formed in the AA collisions. The HRG model also reproduces  \cite {ref13, ref14, ref15} many LQCD calculations on thermodynamic properties of the hadronic phase in the vanishing $\mu_{B}$ region. The topical HRG models do not take into account the effect of finite-size on the system of hadron gas, while matching the yields of AA collisions at varied $\sqrt s_{NN}$ to obtain the freeze-out parameters. Recently, the chemical freeze-out parameters - the temperature, volume and the strangeness saturation factor ($\gamma_{s}$) have been calculated \cite{ref16}, as a function of multiplicity, for pp, pPb and PbPb collisions at $\sqrt s$  = 7, $\sqrt s_{NN}$ = 5.02 and $\sqrt s_{NN}$ = 2.76 TeV, respectively, by matching the measured hadron yields in a common HRG model framework. The study, on the basis of converging results of thermal parameters in different ensembles, concludes that the high-multiplicity events of pp and pA collisions also reach the thermodynamic limit. The conclusion may appear consistent with the observations of collective property of particle production in high-multiplicity pp \cite {ref17, ref18} and pA events \cite {ref19, ref20, ref21} at LHC and RHIC. Intitutively, however, considering the effect of finite system-size becomes important when the hadron gas corresponding to small systems are studied. \\
In this work, to investigate the effect of finite size of hadron gas, in HRG model, we focus on comparison of small systems and large systems, produced in pp collisions at  $\sqrt s$ = 7 TeV and PbPb collisions at $\sqrt s_{NN}$ = 2.76 TeV, respectively. We study the effect on hadron gas systems, in grand canonical ensemble, corresponding to different high multiplicity ($dN_{ch}/d\eta$) pp and PbPb events at the LHC energies and obtain the chemical freeze-out parameters by matching the measured yields of hadrons. We extend the study in terms of thermodynamic variables at the respective freeze-out temperatures, with and without system-size effect, to compare with the ideal thermodynamic system of hadronic phase of strongly interacting matter, provided by the LQCD calculations. The degree of thermalization in these systems has been compared in terms of Knudsen number, also.\\
\section{Yields in HRG model}
\label{}
Measured yields of hadrons in relativistic heavy-ion collisions are matched with calculated yields in HRG-like thermal model, considering a grand canonical ensebmle of hadrons and resonances. In contrast, small yields from small systems of pp collisions at the pre-LHC energies, necessitiated consideration of a canonical ensemble. In this study on high-multiplicity events of pp and AA collisions at the LHC energies, following the recent study \cite {ref16} that describes the hadronic yields in the HRG model for both the systems, we consider grand canonical ensemble for both. Further, as this data driven study focuses on the high-multiplicity events in pp collisions at LHC energies, following the revelation in the study at ref.~\cite {ref16}, we consider the strangeness saturation factor, $\gamma_{s}$ to be unity. \\
The grand canonical partition function for the hadron resonance gas can be written \cite{ref13} as:
\begin{equation} 
\ln Z\textsuperscript{id}=\sum_{i}\pm\frac{Vg_i}{2\pi^2 }\int_{0}^\infty p^2dp\ln \left\{1\pm \exp[(\mu_{i}-E_{i})/T] \right\}
\label{eq:1}
\end{equation}
where $g_{i} $ is the degeneracy factor for $i^{th}$ hadron species, $E_{i}=\sqrt{p^2+m^2_{i}}$ and $\mu_{i} = B_i\mu_B + S_i\mu_s + Q_i\mu_Q $, with the symbols 
carrying their usual meaning. The $(\pm)$ sign corresponds to fermions and bosons, respectively. The pressure $P(T,\mu)$, the energy density $\epsilon(T,\mu)$ and the number density $n(T,\mu)$, obtained from the partition function of all the hadrons, including resonances, and incorporating finite width of the resonances following Breit-Wigner (BW) distribution, are given \cite {ref22, ref23} by: \\
	\begin{eqnarray}
	P(T,\mu) = \pm \frac{T}{2\pi^2 }\sum_i  g_{i}\int_{m^{min}_{i}} ^{m_{i}+2\Gamma_{i}} \rho_{i}^{BW} (m) dm \nonumber\\
	\int_{0}^\infty p^2 dp  \ln(1 {\pm \exp[(\mu_{i} - E_{i})/T])}
	\label{eq:2}
	\end{eqnarray} 
	\begin{eqnarray}
	\epsilon(T,\mu) = \frac{1}{2\pi^2 }\sum_i {g_i}\int_{m^{min}_{i}} ^{m_{i}+2\Gamma_{i}} \rho_{i}^{BW} (m) dm \nonumber\\
	\int_{0}  ^\infty \frac {p^2 dp}  {\exp[(E_i - \mu_i)/T] \pm 1 }E_i
	\label{eq:3}
	\end{eqnarray}
	\begin{eqnarray}
	n(T,\mu) = \frac{1}{2\pi^2 }\sum_i g_{i}\int_{m^{min}_{i}} ^{m_{i}+2\Gamma_{i}} \rho_{i}^{BW} (m) dm \nonumber\\
	\int_{0} ^\infty \frac {p^2 dp}  {\exp[(E_i-\mu_i)/T] \pm 1}
	\label{eq:4}
	\end{eqnarray}
Here, $\rho_{i}^{BW} (m)$, the  Breit-Wigner distribution of $i^{th}$ hadron, is given by :
\begin{equation}
\rho_{i}^{BW} (m)= A_{i}(\frac{2mm_{i}\Gamma_{i}}{(m^{2}-m_{i}^{2})^{2}+m_{i}^{2}\Gamma^{2}_{i}})
\label{eq:5}
\end{equation}
where $A_{i}$ is the normalization constant obtained from $ \int_{m^{min}_{i}} ^{m_{i}+2\Gamma_i} \rho_{i}^{BW} dm$ = 1, ${\Gamma_{i}}$ is the $i^{th}$ resonance width, $m^{min}_{i}$ = $max(m_{i}-2\Gamma_{i},m^{thr}_{i})$ and $m^{thr}_{i} $ represent minimum threshold mass of decay for the $i^{th}$ resonance.	\\
The repulsive interaction among hadrons and resonances is taken care by Excluded Volume (EV) method \cite {ref24} that modifies pressure and chemical potential as:
\begin{equation}
P^{EV}(T,\mu_1,\mu_2,...)=\sum P_i(T,\bar\mu_1,\bar\mu_2,...) 	
\label{eq:6}
\end{equation}
where $\bar\mu_i=\mu_i-v^{EV}_{i}P^{EV}(\mu_1,\mu_2,..)$ and $v^{EV}_{i}=\frac{16}{3} \pi  r_i^3$ is excluded volume of the $i^{th}$ hadron with hard core radius of  $r_{i}$. So, the HRG, 
incorporated with the EV effect (HRGEV), gives the initial or the primordial yield of $i^{th}$ hadron at the chemical freeze-out as: 	
\begin{eqnarray}
N^{prim}_i=\frac{Vn_{i}(T,\bar\mu_i)}{1+\sum_k v^{EV}_{k}n_{k}(T,\bar{\mu_{k}})}
\label{eq:7}
\end{eqnarray}	 \\
The measured total yield of $i^{th}$ hadron, containing the primordial yield and the decay feed down from heavier resonances,
is given by:
\begin{equation} 
N^{total}_{i} = N^{prim}_{i} +\sum_j BR_{j\rightarrow_i}N_{j}
\label{eq:8}
\end{equation}	
where $BR_{j\rightarrow_i}$ is the branching ratio of the decay channel from the $j^{th}$ hadron.\\
Conventionally, the abundances of hadron species are obtained from equation-\ref{eq:8} and fit to the measured yields with $\mu_{B}$, $V$ and $T$ of the hadron gas system as free parameters to find out these parameters ($\mu_{B}^{ch}$, $V^{ch}$ and $T^{ch}$) at the chemical freeze-out. The $\mu_s$ and $\mu_Q$ are extracted from the constraint equations $n_{S}(T,\mu)=0$, for the  strangeness neutrality condition and $n_{Q}(T,\mu)/n_{B}(T,\mu)$ =(Q/B), the ratio of net baryon to net charge of the colliding system \cite {ref25}. \\
\section{System Size effect in MRE}
\label{}
In previous studies \cite {ref26, ref27, ref28} with small systems, the finite-size effect of HRG was implemented by restricting the available phase space through cutting off the low momentum (infrared) regions in the integral over momentum space in the partition function. In this study, we use Multiple Reflection Expansion (MRE) formalism \cite {ref29, ref30, ref31} to implement definite surface that bounds a finite volume of HRG system in grand canonical ensemble, for consideration of the effects due to the surface and its curvature on the density of states of the finite volume. The MRE formalism was first applied to problems in relativistic quantum field theory \cite {ref29} to study the density of states in a cavity, whose boundary can be visualized with reflections from the surface. The MRE formalism has been applied \cite {ref32, ref33} for the partonic phase of strongly interacting matter at zero chemical potential, where the finite-size effect has been found to be operative on systems of radii below 10 $fm$ and around the crossover temperature. Similar formalism, as correction due to finite size, was applied \cite {ref34} while matching thermal model calculations with hadron abundances in AA collisions at SPS-CERN. \\
\begin{figure}[htb!]
	\centering
	\includegraphics[scale=0.40]{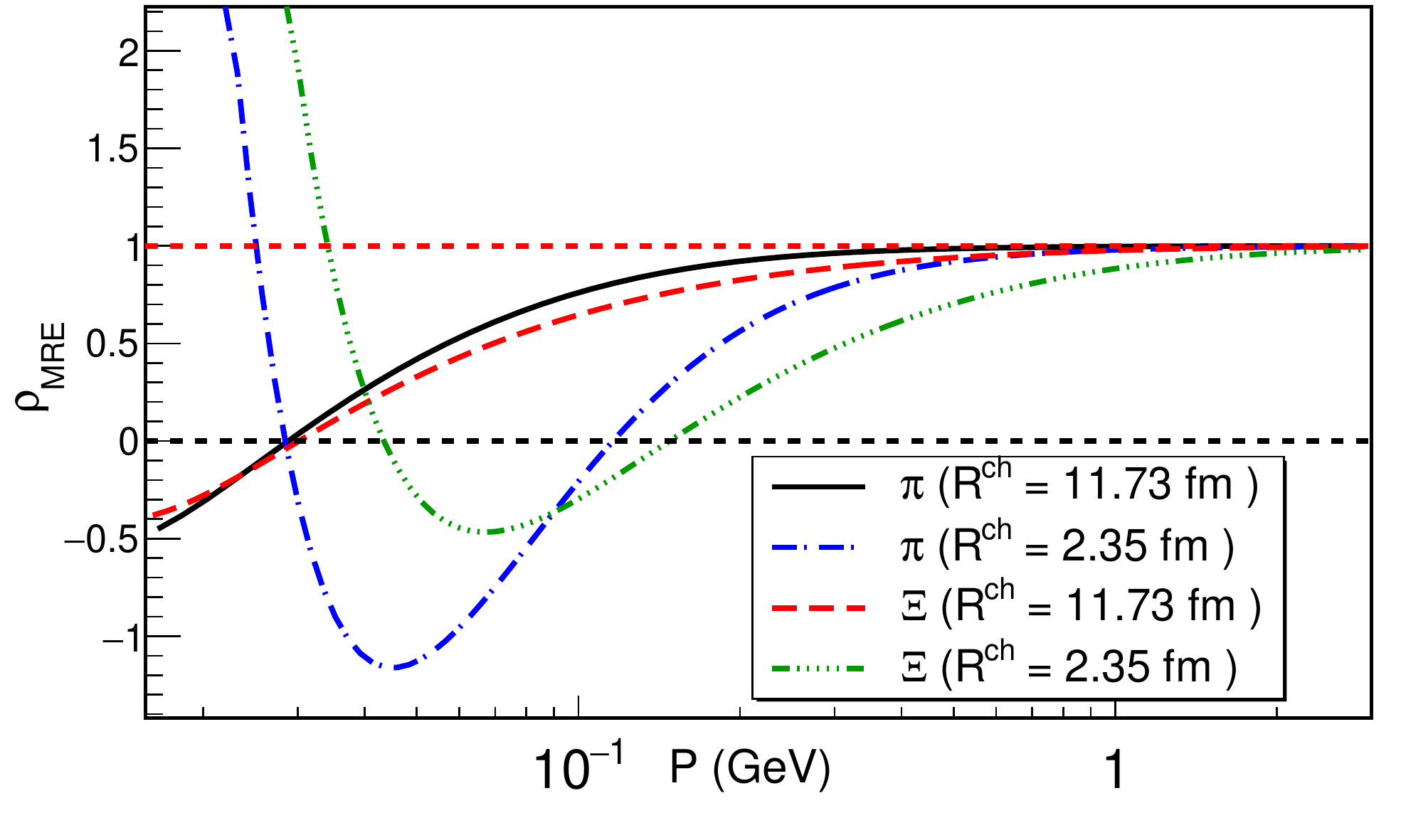}
	\caption{Pictorial presentation of momentum dependence of MRE effect for particles of differnt masses in
		representative system-sizes (corresponding to pp and AA collisions at different centre of mass energy).}
	\label{fig:rho} 
\end{figure}  
The modified grand canonical partition function for the MRE-included HRG model can be written as: 
\begin{eqnarray} 
\ln Z_{mre}^{id}=\sum_{i}\pm\frac{Vg_i}{2\pi^2 } \int_{\lambda_{IR}}^\infty p^2 dp \rho_{i,mre} \nonumber\\
\ln \left\{1\pm \exp[(\mu_{i}-E_{i})/T] \right\}
\label{eq:9} 
\end{eqnarray}
The $\rho_{i,mre}(p,m_i,R)$ gives the density of states, as a function of mass and momentum, modified due to the effect of finite spherical volume of radius R.
\begin{equation}
\rho_{i,mre}(p,m_i,R)= 1 + {\frac { 6\pi^2} {pR}} f_{i,S}(m_i, p) + {\frac {12\pi^2}{({pR})^2}} f_{i,C}(m_i, p) 
\label{eq:10} 
\end{equation}
where the surface contribution is:
\begin{equation}
f_{i,S}(m_i, p) = - {\frac {1} {8\pi}} (1 - {\frac {2} {\pi}}arctan {\frac {p} {m_i}}) 
\label{eq:11} 
\end{equation}
and the curvature contribution is given by Madsen's ansatz \cite {ref30}
\begin{equation}
f_{i,C}(m_i, p) = {\frac {1} {12\pi^2}} [1 - {\frac {3p} {2m_i}} ({\frac {\pi} {2}} -arctan {\frac {p}{m_i}})]
\label{eq:12} 
\end{equation}
As can be understood from equation~\ref{eq:10}, the fraction of the restricted phase space tends to deminish with increasing system-size and / or increasing momentum, taking the value of $\rho_{i,mre}(p,m_i,R)$ tending to unity. The negative values of $\rho_{i,mre}(p,m_i,R)$ that is obtained from the solution of equation $\rho_{i,mre}(p, m_i, R) = 0$ for a given $R$ and $m_{i}$, resulting unphysical density of states, decide the 
IR cut-off $\lambda_{IR}$. In contrast to the finite size effect incorporated in previous studies (\cite{ref26}, \cite{ref27} and \cite{ref28}) with HRG by introducing only system-size dependent lower momentum cutoff, in the MRE formalism, the available phasespace is thus restricted by a mass-dependent lower momentum cutoff $\lambda_{IR}$; for particles of same momentum, the available phase space for higher mass gets more restricted. The effect of MRE can be better understood from the figure~\ref{fig:rho}, where the momentum dependence of MRE effect for particle species of different masses in representative system sizes are presented in figure~\ref{fig:rho}. As can be seen in the figure~\ref{fig:rho}, the mass dependence of the MRE-effect becomes more prominant in small system and for heavier particles.\\
\section{Results and Discussion}
\begin{figure}[htb!]
	\centering
	\includegraphics[scale=0.40]{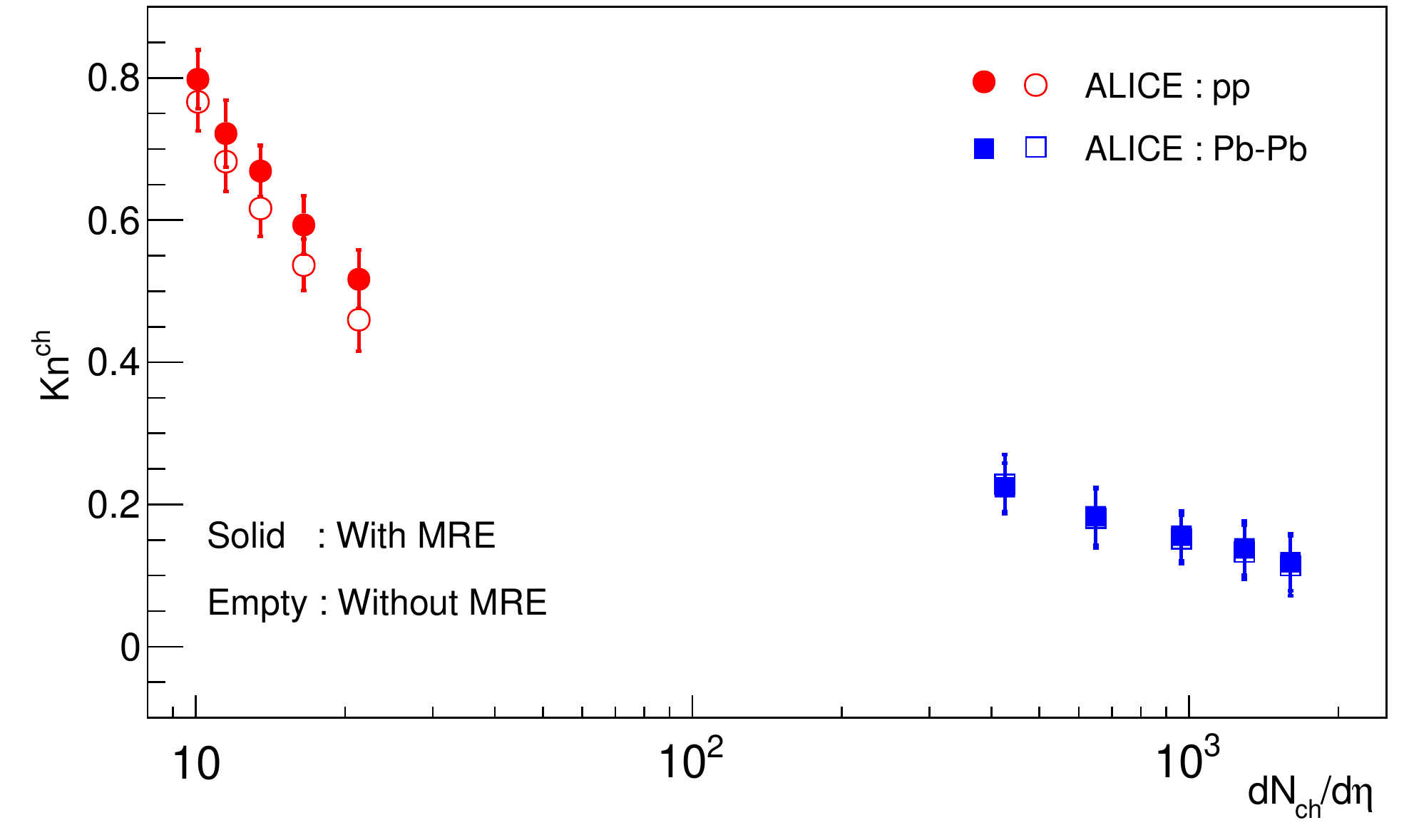}
	\caption{The Knudsen number of hadron gas at chemical freeze-out temperature, with and without MRE effect, corresponding to varying $dN_{ch}/d\eta$ of high-multiplicity events of pp collisions at $\sqrt s$ = 7 TeV  \cite{ref35} and PbPb collisions at $\sqrt s_{NN}$ = 2.76 TeV \cite{ref36, ref37, ref38}.}
	\label{fig:NKn} 
\end{figure}

We obtain chemical freeze-out parameters for several systems of high-multiplicity events of pp ($10.1 \textless dN_{ch}/d\eta \textless 21.3$) collisions at $\sqrt s$ = 7 TeV \cite{ref35} and PbPb ($426 \textless dN_{ch}/d\eta \textless 1601$) collisions at $\sqrt s_{NN}$ = 2.76 TeV \cite{ref36, ref37, ref38} in the HRG model, including i) EV effect and ii) EV + MRE effects, by matching measured yields of hadrons, using the mass table given in Ref. \cite {ref39}. The detail of the optimization of the EV effect can be found in reference~\cite {ref27}. Sometimes, the vanishing $\mu_{B}$ at the LHC energies is approximated to be zero. In our study, we avoid the approximation and use $\mu_{B}$ also as varying parameter. It is important to note that different sets of hadron-species from the same experiment, while fitting the thermal model calculations, result into different sets of chemical freeze-out parameters \cite {ref40}. We consider single freeze-out scenario and extract the freeze-out parameters by simultaneously matching measured yields of charged $\pi$, $K$, $p$, $\Lambda$ and $\Xi$. Obtained freeze-out parameters and corresponding $\chi^2$/ndf, the measure of goodness of fits, from calculations, without and with MRE effects, are tabulated in Tables - I and II, respectively.\\
\begin{table}
\begin{center}
\small\addtolength{\tabcolsep}{-1.3pt}
\begin{tabular}{ |p{1.0cm}| p{1.0cm}|p{1.9cm}|p{1.67cm}| p{1.7cm}| p{1.0cm}|  }
\hline
System &\centering $ \frac{dN_{ch}}{d\eta} $ &\centering $T^{ch}$  {\\ (MeV)}   &\centering $\mu_{B}^{ch}$ {\\ (MeV)} & \centering $R^{ch}$ {\\ (fm)}   & $\chi^{2}/ndf$ \tabularnewline
\hline
\centering PbPb & \centering 1601 & \centering 155.16 (2.18)  & \centering 1.20 (0.09) & \centering 11.68 (0.24) & \centering 2.22\tabularnewline	
\centering PbPb & \centering 1294  & \centering 154.81 (2.02)  & \centering 1.54 (0.16) & \centering 10.74 (0.22) & \centering 2.14\tabularnewline
\centering PbPb & \centering 966 &  \centering 153.77 (2.31)  & \centering 1.16 (0.12) & \centering 9.86 (0.20) & \centering 2.30\tabularnewline
\centering PbPb & \centering 649  &  \centering 153.30 (2.45)  & \centering 1.84 (0.15) & \centering 8.65 (0.22) & \centering 2.85\tabularnewline
\centering PbPb & \centering 426 &  \centering 150.14 (2.38)  & \centering 1.36 (0.12) & \centering 7.73 (0.19) & \centering 2.36\tabularnewline
\centering pp & \centering 21.3   & \centering 157.02 (2.60)  & \centering 0.08 (0.04) & \centering 2.61 (0.08) & \centering 2.51\tabularnewline
\centering pp & \centering 16.5 & \centering 155.29 (2.54)  & \centering 0.11 (0.03)  & \centering 2.52 (0.09) & \centering 3.06\tabularnewline
\centering pp & \centering 13.5 & \centering 153.66 (2.91)  & \centering 0.10 (0.04)  & \centering 2.39 (0.05) & \centering 2.89\tabularnewline
\centering pp & \centering 11.5 & \centering 152.68 (2.45)  & \centering 0.09 (0.03)  & \centering 2.32 (0.06) & \centering 3.76\tabularnewline	
\centering pp & \centering 10.1 & \centering 151.05 (2.61)  & \centering 0.08 (0.03)  & \centering 2.23 (0.06) & \centering 4.34\tabularnewline
\hline
\end{tabular}
\caption{The chemical freeze-out parameters $T_{ch}$, $\mu_{B}^{ch}$ and $R^{ch}$ along with the $\chi^{2}/ndf$, as obtained in HRG model calculations, with the EV effect, by matching the measured hadron yields from events of varied $dN_{ch}/d\eta$ from pp collisions at $\sqrt s_{NN}$ = 7 TeV and PbPb collisions at $\sqrt s_{NN}$ = 2.76 TeV. }
\end{center}
\label{tab:ev}
\end{table}
\begin{table}
\begin{center}
\small\addtolength{\tabcolsep}{-1.3pt}
\begin{tabular}{ |p{1.0cm}| p{1.0cm}|p{1.9cm}|p{1.67cm}| p{1.7cm}| p{1.0cm}|  }
\hline
System &\centering $ \frac{dN_{ch}}{d\eta} $ &\centering $T^{ch}$  {\\ (MeV)}   &\centering $\mu_{B}^{ch}$ {\\ (MeV)} & \centering $R^{ch}$ {\\ (fm)}   & $\chi^{2}/ndf$ \tabularnewline
\hline
\centering PbPb & \centering 1601 & \centering 155.40 (2.10)  & \centering 1.37 (0.12) & \centering 11.73 (0.25) & \centering 2.24\tabularnewline
\centering PbPb & \centering 1294  & \centering 155.24 (2.46)  & \centering 1.55 (0.15) & \centering 10.81 (0.18) & \centering 2.08\tabularnewline
\centering PbPb & \centering 966 &  \centering 154.41 (2.52)  & \centering 1.17 (0.08) & \centering 9.98 (0.21) & \centering 2.28\tabularnewline
\centering PbPb & \centering 649  &  \centering 153.82 (1.94)  & \centering 1.80 (0.12) & \centering 8.75 (0.20) & \centering 2.42\tabularnewline
\centering PbPb & \centering 426 &  \centering 150.84 (2.17)  & \centering 1.05 (0.14) & \centering 7.84 (0.24) & \centering 2.35\tabularnewline
\centering pp  & \centering 21.3   & \centering 157.95 (2.42)  & \centering 0.10 (0.04) & \centering 2.73 (0.07) & \centering 2.33\tabularnewline
\centering pp & \centering 16.5 & \centering 156.58 (2.93)  & \centering 0.09 (0.03)  & \centering 2.62 (0.05) & \centering 2.86\tabularnewline
\centering pp & \centering 13.5 & \centering 155.70 (2.55)  & \centering 0.08 (0.04)  & \centering 2.50 (0.06) & \centering 2.82\tabularnewline
\centering pp & \centering 11.5 & \centering 155.08 (2.15)  & \centering 0.08 (0.04)  & \centering 2.43 (0.07) & \centering 3.22\tabularnewline
\centering pp & \centering 10.1 & \centering 153.76 (2.80)  & \centering 0.09 (0.04)  & \centering 2.35 (0.07) & \centering 3.52\tabularnewline
\hline
\end{tabular}
\caption{Same as Table-I, including the finite-size effect implemented with MRE.}
\end{center}
\label{tab:mre}
\end{table}
The tabulated values of the chemical freeze-out parameters, the control variables of the studied systems in grand canonical ensemble, do not reflect significant effect of the finite system-size. The $\mu_{B}$, in any case approaches zero, as expected. However, as the finite-size effect in MRE formalism is implemented by restricting the density of states, one expects the dependent thermodynamic variables to be sensitive to the effect, as has found in reference~\cite{ref33} in the study of the finite volume effect in the same formalism on partonic medium around the crossover temperature. We study the effect of finite-size on systems of hadron gas at the chemical freeze-out temperature of pp events, characterized by $dN_{ch}/d\eta$, in terms of the pressure and the energy-density.\\   
\begin{figure}[htb!]
\centering
\includegraphics[scale=0.40]{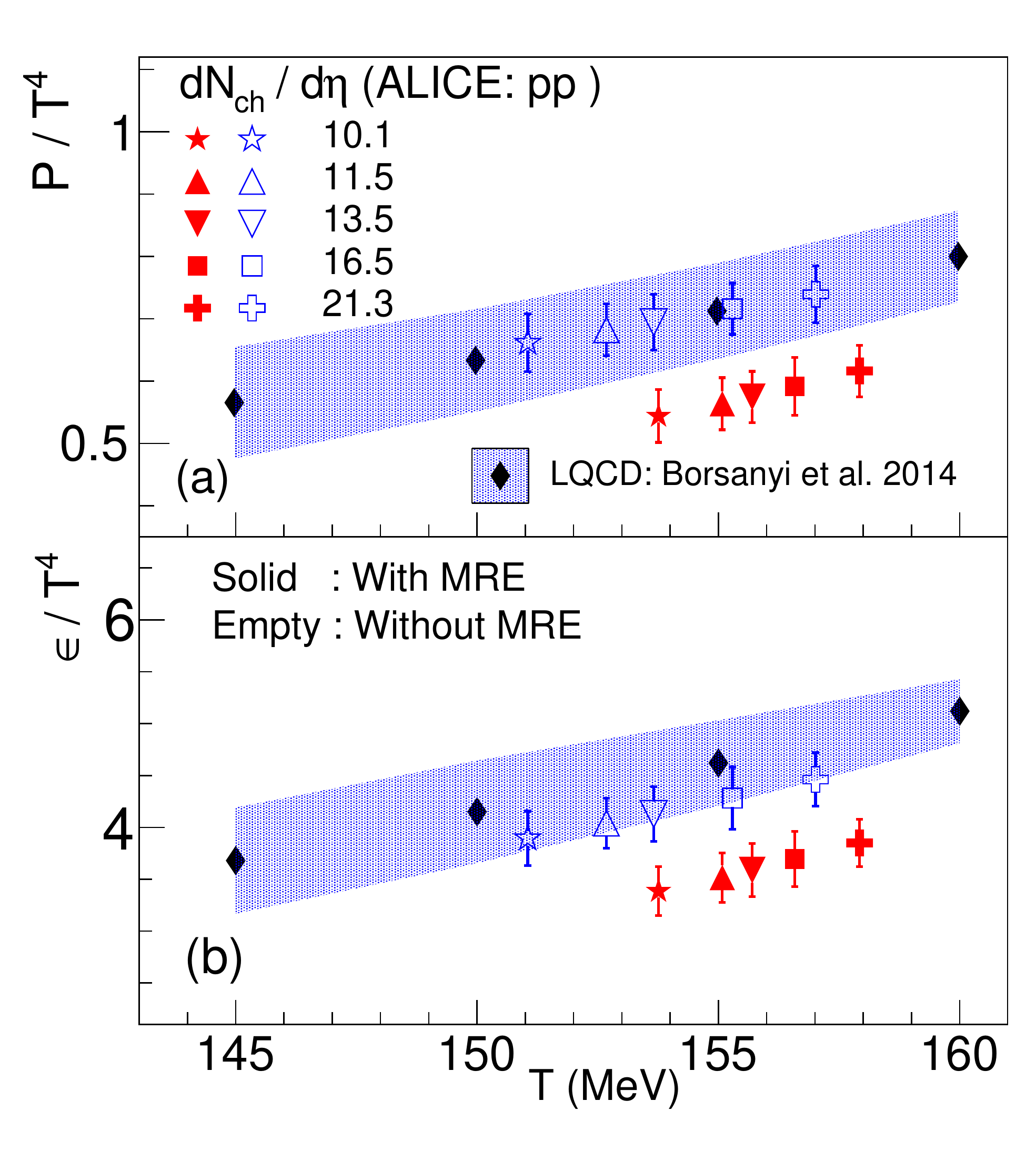}
\caption{Pressure and energy density at chemical freeze-out temperature of hadron gas systems, with and without MRE effect, corresponding to varying $dN_{ch}/d\eta$ of high-multiplicity events of pp collisions \cite{ref35} at $\sqrt s$ = 7 TeV are compared with LQCD results \cite{ref41} (the bands represent the range of uncertainty) on temperature dependence of the variables in panels (a) and (b), respectively.}
\label{fig:PE_pp} 
\end{figure}  
The HRG model calculations, with and without the system-size effect, are compared with the temperature dependence of the thermodynamic variables for hadronic phase of strongly interacting matter, obtained from the the LQCD \cite {ref41} calculation of at $\mu_{B}$ = 0. The comparison allows assessing the degree of thermalization reached in the considered systems with reference to the one at ideal thermodynamic limit. The plots in figure~\ref{fig:PE_pp} show that according to the conventional HRG model formalism (without system-size effect), the thermodynamic limit is apparently reached in hadron gas at the chemical freeze-out temperature of high-multiplicity pp events, approximating the conclusion reached in reference~\cite{ref16} on the basis of equivalence of the freeze-out temperature, the strangeness saturation factor and the volume of the system in different statistical ensembles. However, in contrast to the conclusion in reference~\cite{ref16}, this study, by taking the finite-size effect into account, reveals that the thermodynamic limit is not yet reached for available high-multiplicity pp events of widely varied range of $dN_{ch}/d\eta$ ($10.1 \textless dN_{ch}/d\eta \textless 21.3$). The unexpected matching of results for pp events of wide range of $dN_{ch}/d\eta$ in conventional HRG formalism with those for infinite hadronic system in LQCD may be attributed to the insubstantial application of the HRG model for extracting the chemical freeze-out parameters for thermodynamically infinite system in heavy-ion collisions, to the pp events, without considering the effect of finite size and by judging the goodness of fits in terms of $\chi^{2}/ndf$, that detoriates with decreasing $dN_{ch}/d\eta$. On implementation of the finite-size effect, the values of the thermodynamic variables for pp events with all the considered $dN_{ch}/d\eta$ stay off the ideal condition of thermodynamic limit of hadron gas. The diminishing difference in calculated values of thermodynamic variables, with and without MRE effect and for same $dN_{ch}/d\eta$, with increasing $dN_{ch}/d\eta$ reveals the importance of finite-size effect on small system of hadron gas. The figure~\ref{fig:PE_pbpb}, presenting similar study with the PbPb systems of varied $dN_{ch}/d\eta$, clearly shows that the hadron gas systems, representing different high $dN_{ch}/d\eta$ of PbPb collisions, convincingly reach the thermodynamic limit and the systems remain insensitive to the MRE effect. These plots reiterate the significance of considering the finite-size effect in studying small systems in HRG model.  \\
\begin{figure}[htb!]
	\centering
	\includegraphics[scale=0.40]{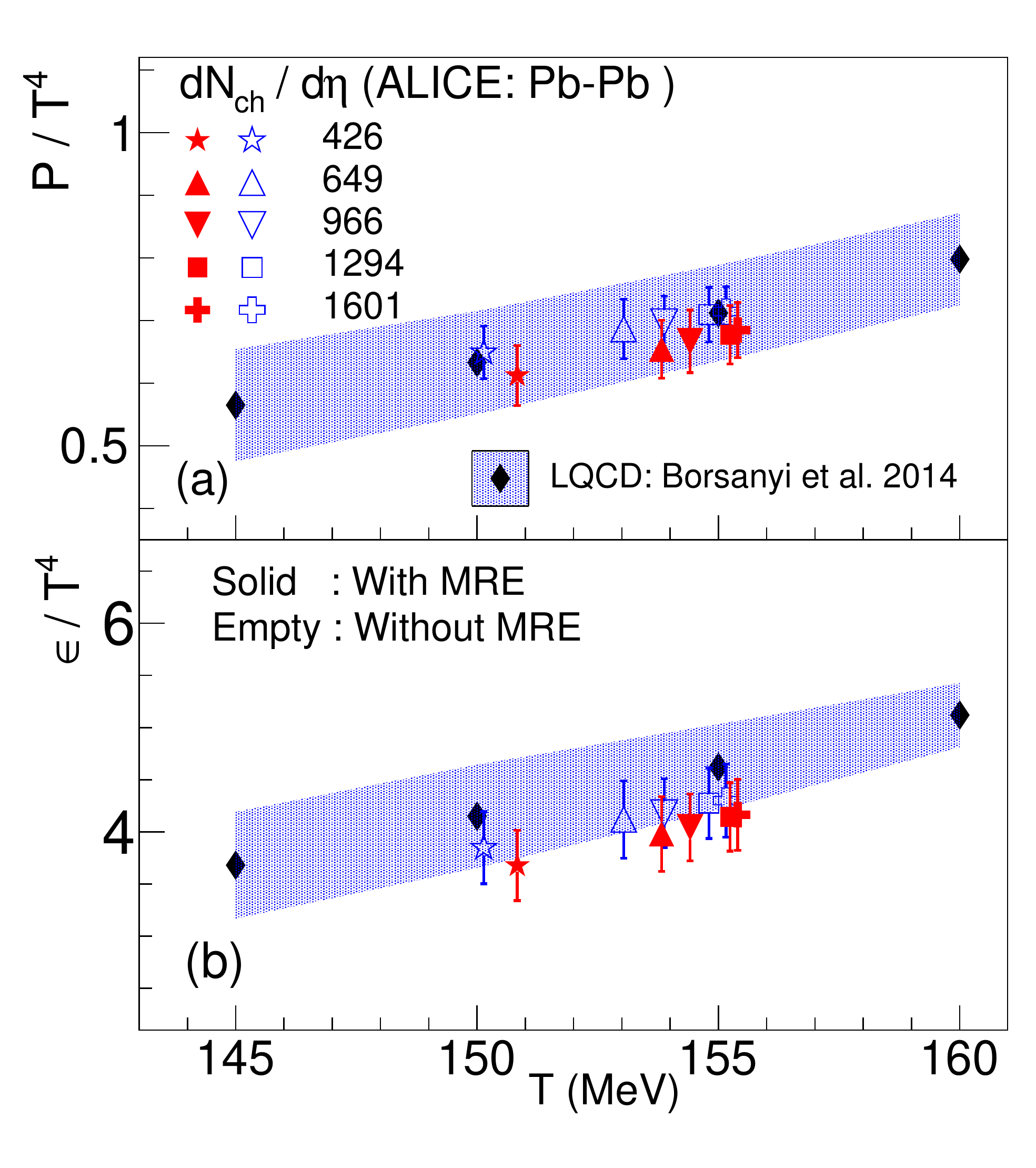}
	\caption{Same as figure~\ref{fig:PE_pp} for PbPb collisions \cite{ref36, ref37, ref38}.}
	\label{fig:PE_pbpb} 
\end{figure}  
We extend the study to the comparison of the degree of thermalization in terms of dimensionless Knudsen number of hadron gas at chemical freeze-out, given by $Kn^{ch} = {{\lambda} / {R^{ch}}}$, where $\lambda^{ch} = 1 / ({n^{ch}}{\sigma})$ is the mean free path, $\sigma$ is the cross-section and $n^{ch}$ is the number density at the corresponding freeze-out temperature. A very small value of Knudsen number indicates that the system of gas is in continiuum region. In a transition region, the continiuum characteristics of the fluid progressively break down with the decreasing size of the system, resulting increase in Knudsen number, till the free streaming particle motion, characterized with a large value of Knudsen number, sets in. It is important to discuss, at this point, the Knudsen number criterion for applicability of fluid dynamics in heavy-ion collisions. It has been suggested  by several \cite {ref42, ref43, ref44, ref45, ref46} hydrodynamic and transport model calculations on Knudsen number for central and semi-central AuAu events at RHIC energy $\sqrt s_{NN}$ = 200 GeV that the criterion for applicability of the fluid dynamics in heavy-ion collisions is $Kn \textless$ 0.5. In a detailed study \cite{ref47} on evoluation of Knudsen number in fluid dynamical modelling of relativistic heavy-ion collisions, the number has been shown \cite{ref47} to have dependence on initialization time and the ratio of shear viscocity to entropy ($\eta/s$). Generally, however, the value of Knudsen number $\textless$ 0.5 fulfills the fluid-dynamical criterion and a value greater than unity corresponds to free-streaming of particles. The condition, 0.5 $\textless$ Kn $\textless$ 1.0 refers to the transition region between the two. We calculate the Knudsen number of hadron gas at chemical freeze-out, for the considered collisions systems. Pions being the most abandunt constituent of the hadron gas systems we use the temperature dependent pion-pion cross-section, $\sigma$, as prescribed in ref.~\cite {ref48} and obtain the number-density from the HRG model, with and without MRE effect. The values of $Kn^{ch}$, as presented in figure~\ref{fig:NKn}, confirms that the degree of thermalization in small and large systems of hadron gas remains far apart. The figure~\ref{fig:NKn} also shows that in contrast to the large system of hadron gas in PbPb events, the small systems corresponding to high-multiplicity pp events have an effect of finite-size consideration. While the PbPb events clearly fulfill the Knudsen number criterion for applicability of fluid dynamics, the $Kn^{ch}$-values for the high-multiplicity pp events, with the system-size effect taken into consideration, lie in the transition region, confirming that these systems remain away from the thermodynamic limit. \\ 
\section{Summary and Remarks}
\label{}
The study on finite-size effect on hadron gas corresponding to high-multiplicity pp events reveals:  
\begin{itemize}
\item  Consideration of the finite-size effect in HRG model calculations, at $\mu_{B}$=0, is important for small systems, like high-multiplicity pp events. 
\item The effect of finite-size on small system of hadron gas is consitent with the observed finite-size effect on small partonic medium around the crossover temperature in the QCD phase diagram. 
\item  Comparison of thermodynamic variables of hadron gas with finite size effect, corresponding to pp events, with LQCD calculations (figure~\ref{fig:PE_pp}) indicates that the thermodynamic limit is not reached in the pp events of avaiable range of high-multiplicity. 
\item The indication of incomplete thermalizatioin in high-multiplicity pp events in terms of Knudsen number is consistent with the conclusion reached from the comparison of thermodynamic variables with the LQCD calculations.
\item The value of Knudsen number decreases with increasing event multiplicity and its value in the trasition region for the highest multiplicity pp events, studied here, is close to the criterion for the applicability of hydrodynamics in heavy-ion collisions. 
\end{itemize}
The decreasing trend of the ($dN_{ch}/d\eta$)-dependent Knudsen number (figure~\ref{fig:NKn}) in the transition region indicates that for pp events of multiplicity higher than the available ones, the Knudsen number may reach the range of applicability of fluid dynamics. The reducing effect of finite size with increasing $dN_{ch}/d\eta$ of pp events (figure~\ref{fig:PE_pp}) suggests the possibility of pp events of higher multiplicity to reach the thermodynamic limit.
  
\end{document}